\documentclass{jpsj-suppl}
\usepackage{txfonts} %Please comment out this line unless the txfonts package is availabe in your LaTeX system.
\usepackage{braket}

\title{Measurements of Neutrino Oscillation Angle $\theta_{13}$}

\author{Masahiro \textsc{Kuze}}

\inst{Department of Physics, Tokyo Institute of Technology, Oookayama 2-12-1-H-26, Meguro, Tokyo 152-8551, Japan}

\email{kuze@phys.titech.ac.jp}

\recdate{January 31, 2013}

\abst{
Neutrinos exhibit an interesting phenomenon called 'neutrino oscillation', in which
a neutrino
changes its flavor after traveling
some flight length.  Many experiments measured the mixing angles and
mass differences, but the angle $\theta_{13}$ had been unmeasured due to its
smallness compared to others.
During 2011 and 2012, series of new-generation neutrino experiments reported
positive results in $\theta_{13}$ search, and its value has been determined to
be just below the previous upper limit.  The non-zero result of $\theta_{13}$
is a very good news for future of neutrino physics, since it opens a possibility
of measuring the CP violation phase in the lepton sector.
An introduction to neutrino oscillation and latest experimental results
are presented.
A detail is put on Double Chooz reactor experiment, in which the author is involved.}

\kword{neutrino, oscillation, mixing, reactor}

\begin{document}
\maketitle

\section{Introduction to Neutrino Oscillation}
Neutrino is a very unique particle.  Its mass is much much smaller than any other elementary
particles - neutrino is supposed to be lighter than 1~eV, while the next lightest particle, electron,
has a mass of 511~keV.  Furthermore, it is a neutral lepton, thus subject to only weak interaction
among the forces in Standard Model.  It interacts with matter very very weakly - the mean free path
of a neutrino with an energy of 1~MeV in water is about $10^{15}$~km, that is about 100 light-years.
    
Neutrino also exhibits an interesting quantum effect called neutrino mixing, i.e. the flavor eigenstates
of three generations $\nu_e, \nu_\mu, \nu_\tau$ are not identical to mass eigenstates $\nu_1, \nu_2, \nu_3$ (with masses $m_1, m_2, m_3$) but are superpositions of them:

\begin{eqnarray}
\left(
    \begin{array}{c}
\nu_e\\
\nu_\mu\\
\nu_\tau
    \end{array}
  \right)
&  = &
\left(
    \begin{array}{ccc}
      U_{e1} & U_{e2} & U_{e3} \\
      U_{\mu 1} & U_{\mu 2} & U_{\mu 3} \\
      U_{\tau 1} & U_{\tau 2} & U_{\tau 3} 
    \end{array}
  \right)
\label{e1}
\left(
    \begin{array}{c}
\nu_1\\
\nu_2\\
\nu_3
    \end{array}
  \right)
\nonumber  \\
& = &
  \left(
    \begin{array}{ccc}
      1 & 0 & 0 \\
      0 & c_{23} & s_{23} \\
      0 & -s_{23} & c_{23} 
    \end{array}
  \right)
    \left(
    \begin{array}{ccc}
      c_{13} & 0 & s_{13}e^{-\mathrm i\delta} \\
      0 & 1 & 0 \\
      -s_{13}e^{\mathrm i\delta} & 0 & c_{13} 
    \end{array}
  \right)
    \left(
    \begin{array}{ccc}
      c_{12} & s_{12} & 0 \\
      -s_{12} & c_{12} & 0 \\
      0 & 0 & 1 
    \end{array}
  \right)
\left(
    \begin{array}{c}
\nu_1\\
\nu_2\\
\nu_3
    \end{array}
  \right)
 \\
& &
(c_{ij} \equiv \cos\theta_{ij}, s_{ij} \equiv \sin\theta_{ij}),
\nonumber
\end{eqnarray}
where $U$ is the Pontecorvo-Maki-Nakagawa-Sakata (PMNS) matrix describing the mixing~\cite{Ponte,MNS},
which is a product of rotation matrices with three mixing angles $\theta_{12}, \theta_{23}, \theta_{13}$.
Also one complex phase, $\delta$,  enters in the matrix which causes CP violation.

The time evolution of a free neutrino follows that for each mass eigenstate with its energy $E_i$:
\begin{eqnarray}
\ket{\nu_\alpha(t)} &=& \sum_i U_{\alpha i} \ket{\nu_i(t)} = \sum_i U_{\alpha i} e^{-\mathrm i E_i t} \ket{\nu_i(0)}  \\
& &(\alpha = e, \mu, \tau)  \nonumber.
\end{eqnarray}
Therefore, the probability of observing flavor $\beta$ for initial flavor $\alpha$ after certain time $t$
includes a term with the factor $U_{\beta j}U_{j\alpha}^{-1}U_{\alpha i}U_{i\beta}^{-1}e^{-\mathrm i (E_i - E_j) t}$, which is a 'beat' of de Broglie waves with
slightly different frequencies.  Since in normal experimental cases the neutrino energy or momentum
is much larger than the mass, an approximation holds $E_i = \sqrt{p^2 + m^2_i} \approx
p + \frac{m^2_i}{2p} \approx E + \frac{m^2_i}{2E}$.
Therefore the energy difference corresponds to mass difference, $\Delta m^2_{ij} = m^2_i - m^2_j$.
For a simple two-flavor case without the complex phase $\delta$, the 'survival probability' of
a neutrino is:
\begin{equation}
P(\nu_\alpha \to \nu_\alpha) = 1 - \sin^2 2\theta \sin^2 \left(\frac{\Delta m^2}{4E} L\right).
\end{equation}
That means, if $\theta$ is non-zero and neutrino has mass, a flavor transmutation is predicted.
This is the origin of the 'neutrino oscillation', and the mixing angles $\theta_{ij}$ are also called
oscillation angles.

\section{Determination of Oscillation Parameters}
In experimental units, neutrino oscillation probability (for simplified two-flavor case) is written as:
\begin{equation}
P(\nu_\alpha \to \nu_\beta) = \sin^2 2\theta \sin^2 \left(1.27\Delta m^2~[\mathrm {eV^2}]~\frac{L}{E}~\left[\mathrm {\frac{km}{GeV}\;or\;\frac{m}{MeV}}\right]\right),
\end{equation}
thus the amplitude of the oscillation is given by $\sin^2 2\theta$ and its wavelength
in $L/E$, with $L$ being the flight length, is
$2hc/\Delta m^2$.
The complete oscillation phenomena are described by three oscillation angles $\theta_{12},
\theta_{23}, \theta_{13}$, two mass differences $\Delta m^2_{32} \equiv m^2_3 - m^2_2, 
\Delta m^2_{21} \equiv m^2_2 - m^2_1$ (since $\Delta m^2_{31} = \Delta m^2_{32} + \Delta m^2_{21}$ and is not independent) and one complex phase $\delta$.

As of 2011, the status of measurements on oscillation parameters were as follows~\cite{textb}:
\begin{itemize}
\item{The solar neutrino experiments (Super-Kamiokande, SNO, ...) and long-baseline
reactor neutrino experiment (KamLAND) revealed $\sin^2 2\theta_{12} \approx 0.8,
\Delta m^2_{21} \approx 8 \times 10^{-5} \mathrm {eV^2}$.}
\item{The atmospheric neutrino measurement (Super-Kamiokande) and accelerator
long-baseline experiments (K2K, MINOS) revealed  $\sin^2 2\theta_{23} \approx 1.0,
|\Delta m^2_{32}| \approx 2.5 \times 10^{-3} \mathrm {eV^2}$.}
\item{Only upper limit on $\theta_{13}$ was given by CHOOZ reactor experiment,
$\sin^2 2\theta_{13} < 0.15$, while $\Delta m^2_{31} \approx \Delta m^2_{32}$.}
\item{The CP-violating phase $\delta$ was completely unknown.}
\end{itemize}
Therefore, it was known $\theta_{12} \approx 34^\circ, \theta_{23} \approx 45^\circ$
but $\theta_{13} < 11^\circ$.  Also it was known $|\Delta m^2_{32}| \gg |\Delta m^2_{21}|$
(by a factor of $\sim$ 30) and the sign of $\Delta m^2_{32}$ was unknown (mass hierarchy).

Thus a positive measurement of $\theta_{13}$ was an imminent goal of neutrino physics
during the past few years.  The importance of its measurement was two-fold:
\begin{itemize}
\item{This was the last unmeasured mixing angle, so the measurement itself had a fundamental
value.  The structure of PMNS matrix was very unique - the non-diagonal elements were
fairly large, in contrast to Cabbibo-Kobayashi-Maskawa mixing matrix in quark sector,
which is mostly diagonal.  Clarification of mixing matrices would be a key to the origin
of flavor, i.e. generation of elementary particles.}
\item{To measure the CP-violation phase $\delta$ experimentally, the condition
$\theta_{12}\theta_{23}\theta_{13}\neq 0$ is necessary.  Therefore, without knowing
the value of $\theta_{13}$, it is not possible to plan future large neutrino experiments
to measure the leptonic CP violation.  The leptonic CP violation is considered to be
the key to explain the matter-antimatter asymmetry of the Universe.}
\end{itemize}

\section{Methods of Measuring $\theta_{13}$}
The difficulty of $\theta_{13}$ measurement lies in its smallness compared to the other two angles.
To measure its value, two experimental methods are possible.
One is to look for appearance of $\nu_e$ from a $\nu_\mu$ accelerator beam.
The energy is in $E\sim$ 1~GeV range and thus corresponds to $L$ = 300$\sim$700~km
with $\Delta m^2_{31} (\Delta m^2_{32})$ oscillation.
The detector looks for electrons, and clean signal can be obtained if the background
is low enough.
On the other hand, the oscillation probability includes many parameters other than
$\theta_{13}$, thus extracting its value from the measurement has theoretical complication.
T2K experiment in Japan and MINOS experiment in US are such accelerator experiments.

The other method is to measure the disappearance of electron anti-neutrino ($\bar \nu_e$)
emitted from nuclear reactors.  The energy of detectable neutrino is $E\sim$ 4~MeV.
It undergoes a $\Delta m^2_{21}$ oscillation with a large probability driven by $\theta_{12}$
at a distance of $\sim$ 100~km, which was exactly what KamLAND experiment measured.
However, it also undergoes a $\Delta m^2_{31}$ oscillation at a shorter distance, $L$ = 1$\sim$2~km,
with much smaller $\theta_{13}$.  The energy is too low to generate $\mu$ or $\tau$ leptons
for the transmuted neutrinos, thus only a deficit of electron events can be measured.
To measure the small disappearance, a precise measurement with large statistics and
small systematic uncertainty is critical, but the measurement is directly connected to the
value of $\theta_{13}$ and pure measurement is possible.
Double Chooz in France, Daya Bay in China and RENO in Korea are such new-generation
reactor neutrino experiments dedicated for $\theta_{13}$ measurement.
One should also note that two kinds of experiments, accelerator and reactor, are
complementary and by combining both results further constraints on osillation parameters
can be expected.

The T2K (Tokai-to-Kamioka) experiment uses a muon neutrino beam from J-PARC
accelerator complex in Tokai, Japan.
The neutrinos are detected by Super-Kamiokande detector after traveling 295~km.
In June 2011, T2K first showed an indication of electron neutrino appearance~\cite{T2KPRL}.
It observed six electron events after selection while 1.5$\pm$0.3 background events
were expected, which corresponds to 2.5~$\sigma$ statistical significance.
In summer 2012, T2K showed an updated preliminary result with roughly double
the statistics: 11 events observed with expected background of 3.2$\pm$0.4, corresponding to
$3.2~\sigma$ evidence~\cite{T2Kprel}.

In the next sections, details are described on reactor neutrino experiments, with Double Chooz as an example.
    
\section{Reactor Neutrino Experiment}
As briefly explained in the previous section, the reactor $\theta_{13}$ experiment
looks for a small deficit of electron anti-neutrinos at a distance of $L = 1\sim 2$~km, which
is the oscillation maximum driven by $\Delta m^2_{31}$ for $E\sim$ 4~MeV:
\begin{equation}
P(\bar\nu_e \to \bar\nu_e) = 1 - \sin^2 2\theta_{13} \sin^2 (1.27\Delta m^2_{31}L/E) + \mathcal{O}(10^{-3}).
\end{equation}
The third, non-leading, term is very small at these distances thanks to the smallness of
$\Delta m^2_{21}/\Delta m^2_{31}$, such that the experiment is a pure $\theta_{13}$ measurement.

The principle of the experiment is to place a neutrino detector at a suitable distance from the
reactor core(s) and to count neutrinos.
The reactor neutrino flux and spectrum are calculable and also have been measured by
past experiments.
The rate of interactions at the (far) detector is predictable from the distance, target mass
(number of protons) and the interaction cross section (inverse beta-decay explained later),
and also as a function of the possible neutrino oscillation parameters.
As the oscillation probability is a function of $L/E$, the energy spectrum of the measured
neutrinos also carries information of the oscillation, so precise energy measurement is helpful.

Even better is to place a 'near' detector at a shorter distance,
and measure  {\it in situ} the neutrino flux and spectrum before the oscillation.
This reduces the systematic uncertainties related to the absolute rate prediction,
since most of them cancel by taking a ratio of measurements by the two detectors.
Placing two 'identical' detectors at near and far distances is ideal, and such 'double' detector
strategy is adopted by all modern experiments, Double Chooz, Daya Bay and RENO.

To detect reactor anti-neutrinos, inverse-beta decay reaction is used:
\begin{equation}
\bar\nu_e + p \to e^+ + n.
\end{equation}
The experiment uses liquid scintillator as a target, containing many free protons.
The energy threshold for the reaction is 1.8~MeV.  The positron quickly loses its kinetic energy,
which is $E_\nu - 1.8~\rm MeV$, and annihilates with an electron.  Therefore, together with the annihilation
photons, this 'prompt signal' gives an estimate of the neutrino energy: $E_{\rm prompt} = E_\nu -
1.8~{\rm MeV} + 2m_e = E_\nu - 0.8~\rm MeV.$

The neutron receives a small recoil energy and drifts in the scintillator, and gets thermalized
and captured by gadolinium (Gd) nuclei doped in the scintillator.
Gd has a very large neutron absorption
cross section, and the resulting excited nucleus emits a few photons to go back to the ground
state, giving an energy deposit of about 8~MeV.
This 'delayed signal' is separated from the prompt signal by the neutron thermalization time,
which is 30~$\mu$s on average.
Therefore, by taking a {\it delayed coincidence} of the two signals, one can reduce the natural
background drastically and separate the rare neutrino reaction.
In addition, the characteristic peak of 8~MeV given by the delayed signal is much higher
than the natural radioactivity, helping to reduce the accidental background
(without Gd, the neutron would be captured by the hydrogen yielding a photon of 2.2~MeV).

The cross section of the reaction increases as $\approx E^2$ above the threshold, and with the
falling spectrum of neutrinos from reactor (beta decays of fission products), the detected
neutrinos will peak at an energy of 3.5$\sim$4~MeV.
In the next section, Double Chooz experiment and its results will be described.

\section{Double Chooz Experiment}
The experiment uses two reactors in Chooz Nuclear Power Plant in Chooz village, Ardenne, France.
Each reactor operates nominally at 4.27 GW$_{\rm th}$ (thermal power, which is usually three times the
generated electric power),
yielding about $10^{21}$ $\bar\nu_e$'s every second.
The far detector is located at 1050~m from the reactor cores, and the near detector is being
built at 400~m, to be completed in 2013.
Therefore, the results shown in the following have been obtained using the far detector only,
without benefitting from the reduction of flux uncertainty by making a relative measurement.

The setup of the detector is shown in Fig.~{\ref{f1}.  The innermost part called Inner Detector
is subdivided into three volumes separated by optically transparent cylindrical vessels (made of
acrylics).
The innermost volume is Neutrino Target, filled with liquid scintillator with Gd loading.
The next volume is Gamma Catcher, also liquid scintillator but without Gd.
Then the Buffer volume surrounds, filled with non-scintillating mineral oil.
The Buffer is enclosed in a stainless-steel vessel and 390 photomultipliers (PMTs)
are mounted to detect scintillation light from the inner two volumes.

The Neutrino Target defines the interaction fiducial volume, about 10~$\rm m^3$, tagged with the 8~MeV
peak of the delayed energy.
The Gamma Catcher is designed to measure the $\gamma$ rays from the interaction
occurring near the edge of the Target, thus helping to improve the energy resolution.
The Buffer protects the inner active volumes from the environmental radiation, such as
$\gamma$ rays from PMT material (mainly glass) or fast neutrons from cosmic-ray
interactions outside the detector.

Double Chooz uses Hamamatsu R7081 10-inch PMTs for Inner Detector, especially modified
using low-background glass developed for this experiment.
Surrounding the Buffer is the Inner Veto volume, filled with liquid scintillator and viewed
by different PMTs.  It tags cosmic muons entering the detector.
Finally a 15~cm thick steel shielding protects the main detector from environmental
radiation from the surrounding rocks.

Above the main detector, there are two layers of plastic scintillator strips, called Outer Veto.
It tags and tracks cosmic muons coming from above.
There is also the Glove Box above the chimney of the liquid vessels.
Some calibration devices like radioactive sources or optical elements can be
inserted periodically to check the response of the detector.

    \begin{figure}
        \begin{minipage}[t]{0.45\linewidth}
%            \includegraphics[width=0.95\linewidth]{detector.eps}
%            \caption{The schematic view of the Double Chooz detector.}
            \includegraphics[width=0.95\linewidth]{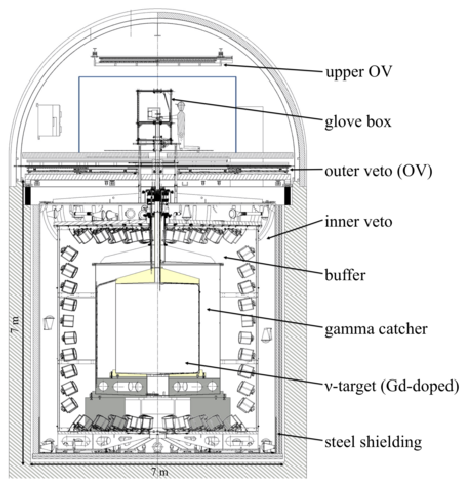}
            \caption{The schematic view of the Double Chooz detector.
            The buffer volume is enclosed in a stainless-steel vessel, which holds 390 PMTs
            (depicted as small boxes).}
            \label{f1}
        \end{minipage}
        \hspace{0.05\linewidth}
        \begin{minipage}[t]{0.45\linewidth}
            \includegraphics[width=0.95\linewidth]{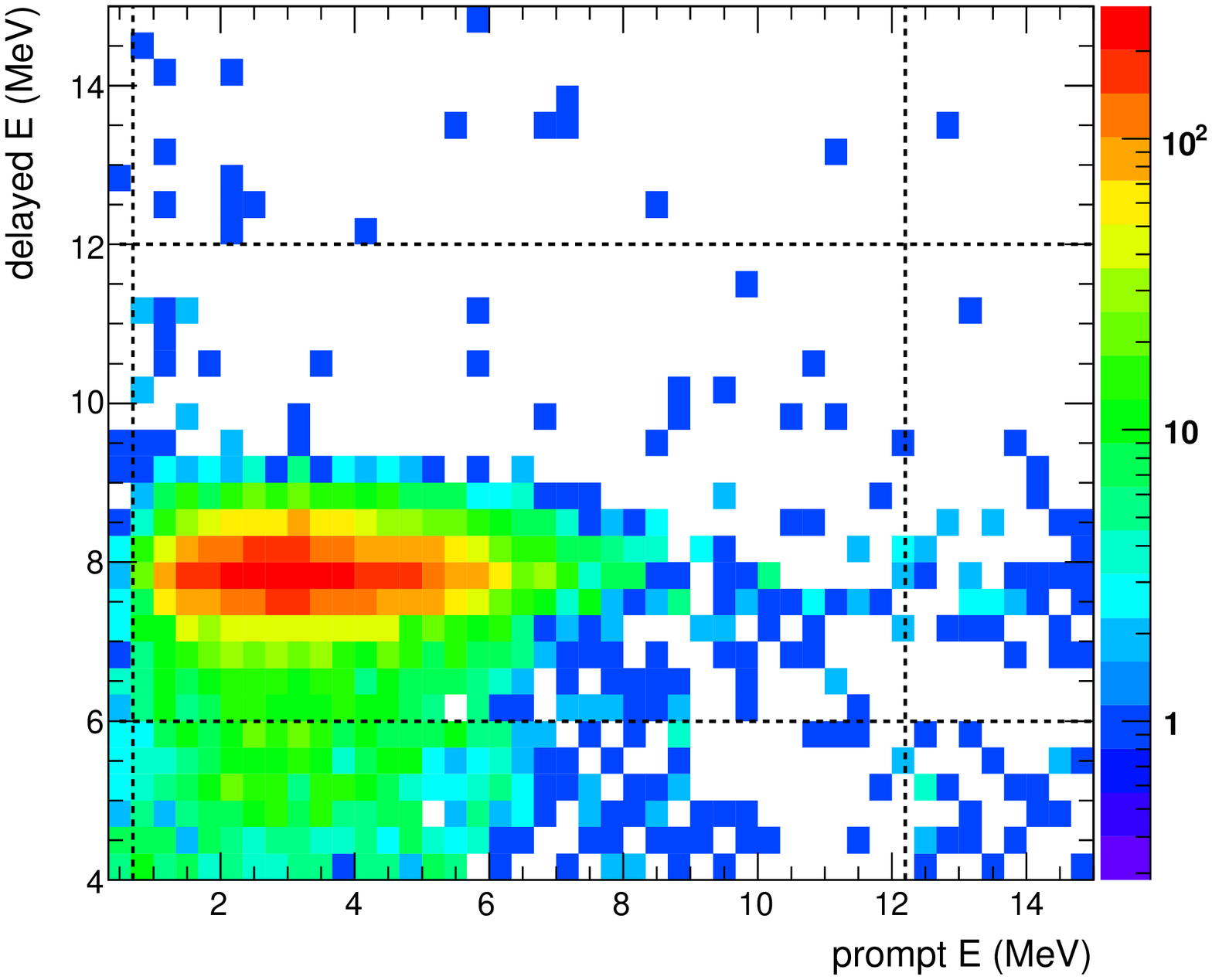}
            \caption{Delayed energy vs. prompt energy for selected neutrino candidates.
            The dashed lines show the energy cuts for selection.}
            \label{f2}
        \end{minipage}
\vspace{-0.1cm}
    \end{figure}

The experiment released its first result in Nov. 2011, with 101 days of data-taking
live-time~\cite{DC1st}.
An indication of deficit was observed, with 94.6~\% confidence level.
Corresponding oscillation parameter was determined to be 
$\sin^2 2\theta_{13} = 0.086 \pm 0.041(\rm stat.) \pm 0.030(\rm syst.)$.
In June 2012, an updated result with 228 days live-time was released~\cite{DC2nd}.
Figure~\ref{f2} shows the distribution of delayed energy vs. prompt energy for the
selected neutrino candidates (time coincidence of 2 to 100 $\mu$s).
A clear peak of delayed energy at 8~MeV is observed, and neutrino events
are selected with small backgrounds (a few~\% level).
Figure~\ref{f3} shows the prompt energy spectrum.  
An oscillation fit yields 
$\sin^2 2\theta_{13} = 0.109 \pm 0.030(\rm stat.) \pm 0.025(\rm syst.)$,
establishing a non-zero $\theta_{13}$ value at 2.9 $\sigma$.

      \begin{figure}
        \begin{minipage}[t]{0.45\linewidth}
            \includegraphics[width=0.95\linewidth]{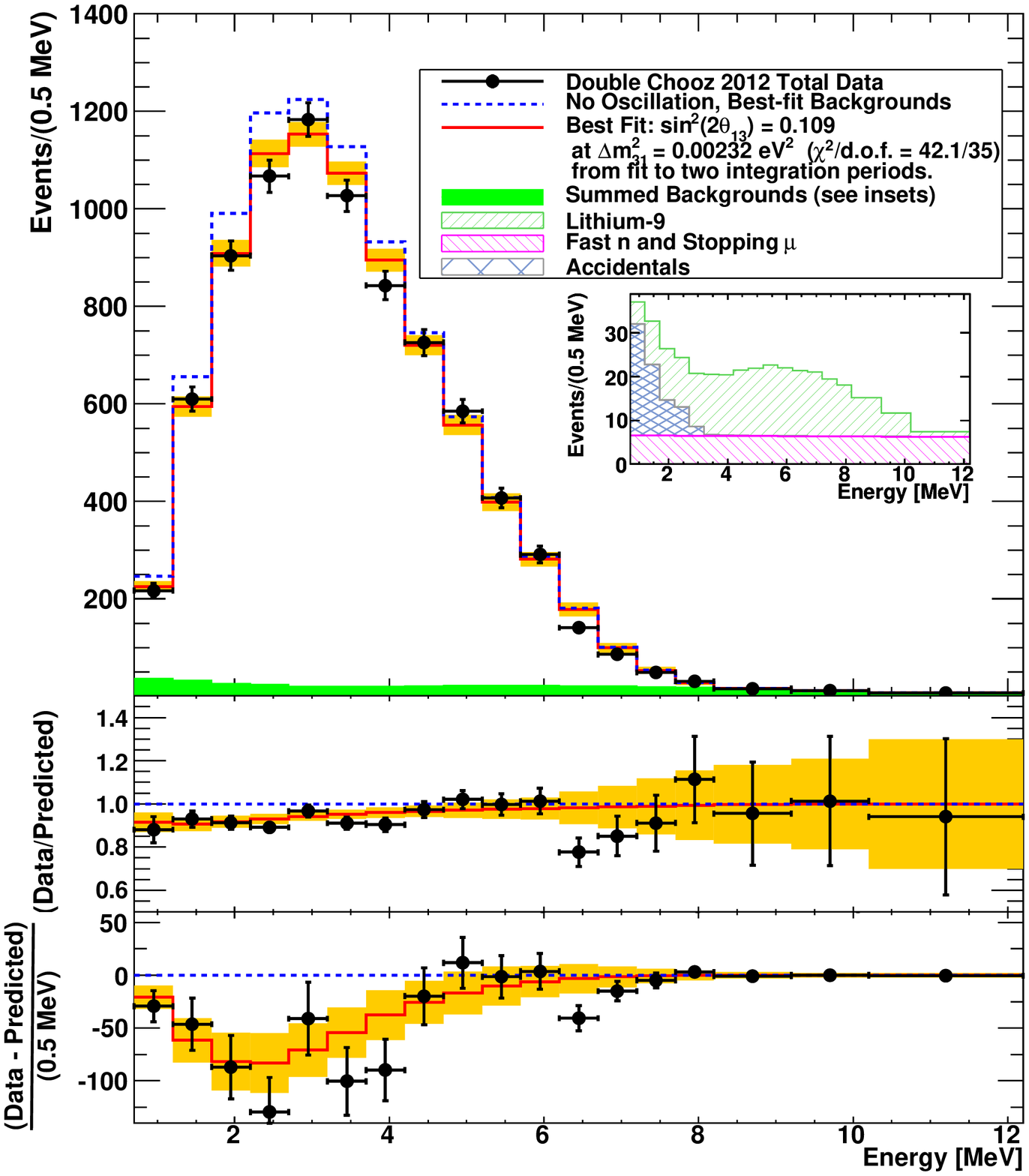}
            \caption{(top) The prompt energy spectrum from Double Chooz data, compared with
            predictions;  (middle) ratio and (bottom) difference of data and prediction without oscillation.}
            \label{f3}
        \end{minipage}
        \hspace{0.05\linewidth}
        \begin{minipage}[t]{0.45\linewidth}
            \includegraphics[width=0.95\linewidth]{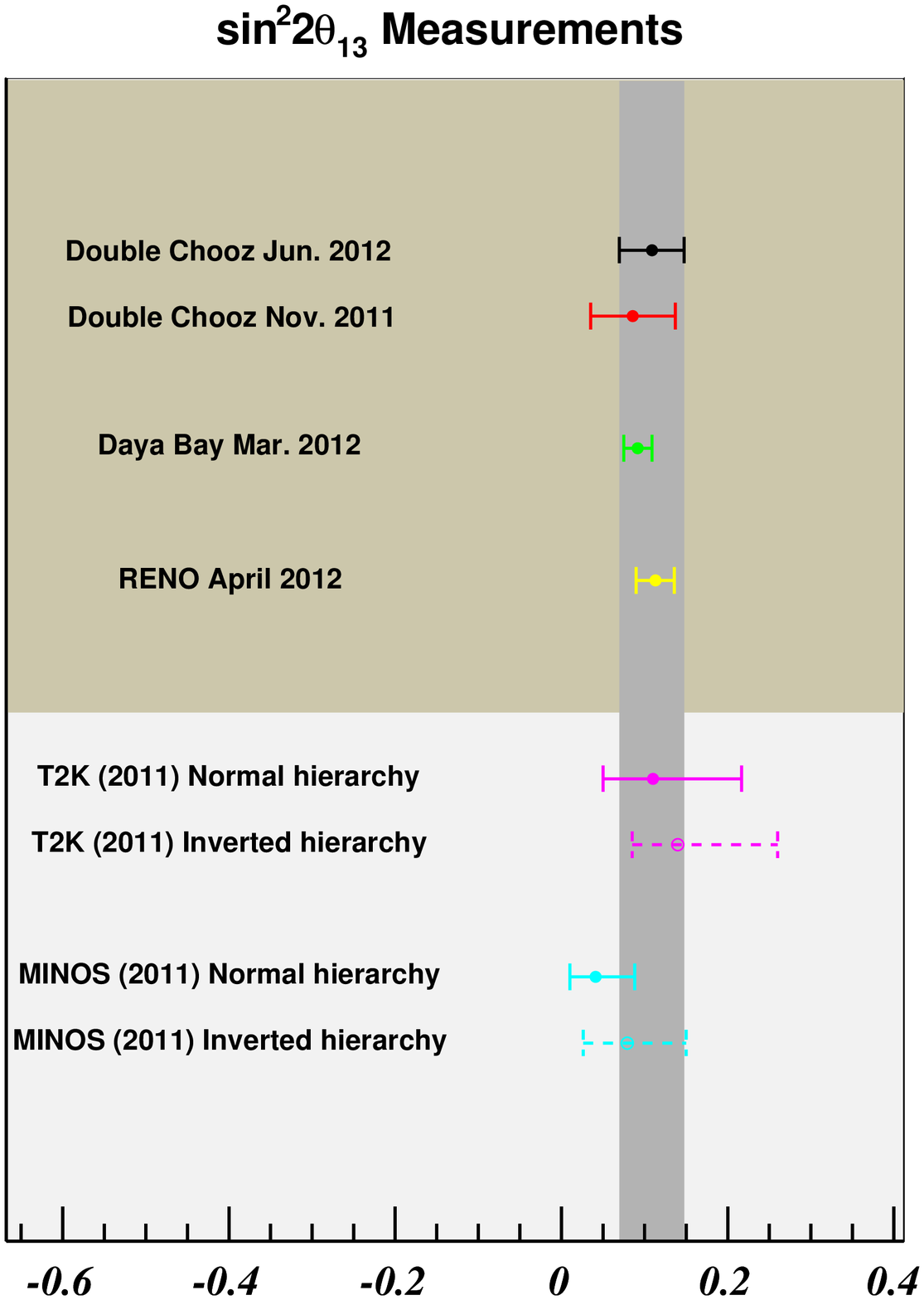}
            \caption{Comparison of $\theta_{13}$ measurements by various experiments.}
            \label{f4}
        \end{minipage}
\vspace{-0.1cm}
    \end{figure}

\section{Status of $\theta_{13}$ Measurements}
Figure~\ref{f4} is from Double Chooz second publication~\cite{DC2nd},
showing a comparison of $\theta_{13}$ measurements.
Daya Bay experiment in China released its first result~\cite{DB1st} in March 2012
with high-statistics data using both near and far detectors.
It yielded $\sin^2 2\theta_{13} = 0.092 \pm 0.016(\rm stat.) \pm 0.005(\rm syst.)$,
an observation with 5.2~$\sigma$.
RENO experiment in Korea~\cite{RENO} also published its result in April 2012
with near and far detectors, yielding
$\sin^2 2\theta_{13} = 0.113 \pm 0.013(\rm stat.) \pm 0.019(\rm syst.)~(4.9~\sigma)$.
Daya Bay just recently published an updated paper~\cite{DB2nd} with more than
2.5 times statistics, with $\sin^2 2\theta_{13} = 0.089 \pm 0.010(\rm stat.) \pm 0.005(\rm syst.)$.

As can be seen, all experiments gave consistent values,  $\theta_{13} \approx 9^\circ$,
which was just below the previous limit.
The $\theta_{13}$ experiments entered into precision measurement era from discovery era.
As stated before, it is good to have both reactor and accelerator experiments, since
combination of measurements could give further constraints on other oscillation parameters.
Double Chooz will also have the near detector in 2013, and with somewhat shorter baseline
than the other two reactor experiments, global analysis of three experiments will be useful
in investigating the oscillation phenomena further.

The relatively large value of $\theta_{13}$ is a very good news for neutrino physics,
since it opens an opportunity of future large-scale experimental programs to discover the CP violation
phase $\delta$ in neutrino sector.
An example is Hyper-Kamiokande proposal\cite{HK}, a water \v{C}erenkov detector
20-times larger than Super-Kamiokande, together with high-power neutrino beam from upgraded J-PARC.

\section{Summary}
Neutrino is a very unique particle.  
It exhibits an interesting quantum phenomenon, neutrino oscillation.
Many experiments had revealed the oscillation parameters, but the angle $\theta_{13}$ was unmeasured since it is much smaller than the other two.
During 2011-12, series of reactor and accelerator experiments have finally established the non-zero value of $\theta_{13}$, which was around $9^\circ$.
Its relatively large value is a very good news for future experiments to further explore CP violation in lepton
sector, which could be a key for matter-antimatter asymmetry of the Universe.
%You can use this file as a template to prepare your manuscript for Supplement to the Journal of the Physical Society of Japan\cite{jpsj,instructions}.
%
%Copy \verb|jpsj-suppl.cls| and \verb|cite.sty| onto an arbitrary directory under the texmf tree, for example, \verb|$texmf/tex/latex/jpsj|. If you have already obtained \verb|cite.sty|, you do not need to copy it.

%\begin{table}[tbh]
%\caption{Captions to tables and figures should be sentences.}
%\label{t1}
%\begin{tabular}{ll}
%\hline
%AAA & BBB \\
%CCC & DDD \\
%\hline
%\end{tabular}
%\end{table}

\end{document}